\documentclass[]{aastex631}
\usepackage{amsmath,amssymb,amsfonts}%

\begin{document}

\title{Observation of the Forbush decrease on 2024 May 10, using the ALPAQUITA air-shower array at the 70-1000\,GV rigidity range}
\shorttitle{Forbush decrease observed by ALPAQUITA}

\author[0000-0002-7415-3611]{M.~Anzorena}
\affiliation{Institute for Cosmic Ray Research, 277-8582, Kashiwa, Japan}
\author[0000-0001-9643-4134]{E.~de~la~Fuente}
\affiliation{Institute for Cosmic Ray Research, 277-8582, Kashiwa, Japan}
\affiliation{Departamento de Fisica, CUCEI, Universidad de Guadalajara, Blvd. Marcelino Garc\'{i}a Barrag\'{a}n \#1421, esq Calzada Ol\'{i}mpica, C.P., 44430, Guadalajara, Jalisco,Mexico}
\affiliation{Doctorado en Tecnolog\'{i}as de la Informaci\'{o}n, CUCEA, Universidad de Guadalajara, Zapopan, M\'{e}xico}
\author[0009-0006-8756-3333]{K.~Fujita}
\affiliation{Institute for Cosmic Ray Research, 277-8582, Kashiwa, Japan}
\author[0000-0002-1172-8289]{R.~Garcia}
\affiliation{Institute for Cosmic Ray Research, 277-8582, Kashiwa, Japan}
\author[0000-0002-0890-0607]{Y.~Hayashi}
\affiliation{Department of Science and Technology, Shinshu University, 390-8621, Matsumoto, Nagano, Japan}
\author[0000-0001-9259-6371]{K.~Hibino}
\affiliation{Faculty of Engineering, Kanagawa University, 221-8686, Yokohoma, Kanagawa, Japan}
\author{N.~Hotta}
\affiliation{Utsunomiya University, 321-8505, Utsunomiya, Tochigi, Japan}
\author{G.~Imaizumi}
\affiliation{Institute for Cosmic Ray Research, 277-8582, Kashiwa, Japan}
\author{Y.~Katayose}
\affiliation{Faculty of Engineering, Yokohama National University, 240-8501, Yokohama, Kanagawa, Japan}
\author[0000-0002-4913-8225]{C.~Kato}
\affiliation{Department of Physics, Shinshu University, 390-8621, Matsumoto, Nagano, Japan}
\author{S.~Kato}
\affiliation{Institut d’Astrophysique de Paris, CNRS UMR 7095, Sorbonne Universit\'{e}, 98 bis bd Arago 75014, Paris, France}
\author{T.~Kawashima}
\affiliation{Institute for Cosmic Ray Research, 277-8582, Kashiwa, Japan}
\author{K.~Kawata}
\affiliation{Institute for Cosmic Ray Research, 277-8582, Kashiwa, Japan}
\author{M.~Kobayashi}
\affiliation{Faculty of Engineering, Yokohama National University, 240-8501, Yokohama, Kanagawa, Japan}
\author{S.~Kobayashi}
\affiliation{Department of Physics, Shinshu University, 390-8621, Matsumoto, Nagano, Japan}
\author{T.~Koi}
\affiliation{College of Engineering, Chubu University, 487-8501, Kasugai, Aichi, Japan}
\author{H.~Kojima}
\affiliation{Chubu Innovative Astronomical Observatory, Chubu University, 487-8501, Kasugai, Aichi, Japan}
\author{P.~Miranda}
\affiliation{Instituto de Investigaciones F\'{i}sicas, Universidad Mayor de San Andres, 8635, La Paz, Bolivia}
\author{S.~Mitsuishi}
\affiliation{Department of Physics, Shinshu University, 390-8621, Matsumoto, Nagano, Japan}
\author{A.~Mizuno}
\affiliation{Institute for Cosmic Ray Research, 277-8582, Kashiwa, Japan}
\author[0000-0002-2131-4100]{K.~Munakata}
\affiliation{Department of Physics, Shinshu University, 390-8621, Matsumoto, Nagano, Japan}
\author{Y.~Nakamura}
\affiliation{Institute for Cosmic Ray Research, 277-8582, Kashiwa, Japan}
\author{M.~Nishizawa}
\affiliation{National Institute of Informatics, 101-8430, Chiyoda, Tokyo, Japan}
\author{Y.~Noguchi}
\affiliation{Faculty of Engineering, Yokohama National University, 240-8501, Yokohama, Kanagawa, Japan}
\author{S.~Ogio}
\affiliation{Institute for Cosmic Ray Research, 277-8582, Kashiwa, Japan}
\author[0000-0002-5056-0968]{M.~Ohishi}
\affiliation{Institute for Cosmic Ray Research, 277-8582, Kashiwa, Japan}
\author{M.~Ohnishi}
\affiliation{Institute for Cosmic Ray Research, 277-8582, Kashiwa, Japan}
\author{A.~Oshima}
\affiliation{Graduate School of Engineering, Chubu University, 487-8501, Kasugai, Aichi, Japan}
\affiliation{College of Engineering, Chubu University, 487-8501, Kasugai, Aichi, Japan}
\author{M.~Raljevic}
\affiliation{Instituto de Investigaciones F\'{i}sicas, Universidad Mayor de San Andres, 8635, La Paz, Bolivia}
\author{H.~Rivera}
\affiliation{Instituto de Investigaciones F\'{ı}sicas, Universidad Mayor de San Andres, 8635, La Paz, Bolivia}
\author{T.~Saito}
\affiliation{Tokyo Metropolitan College of Industrial Technology, 116-8523, Arakawa, Tokyo, Japan}
\author[0000-0002-2589-8561]{T. Sako}
\affiliation{Institute for Cosmic Ray Research, 277-8582, Kashiwa, Japan}
\author[0000-0002-7003-6493]{T.~K.~Sako}
\affiliation{Department of Information and Electronics, Nagano Prefectural Institute of Technology, 386-1211, Ueda, Nagano, Japan}
\author[0000-0003-1462-6365]{S.~Shibata}
\affiliation{Chubu Innovative Astronomical Observatory, Chubu University, 487-8501, Kasugai, Aichi, Japan}
\author{A.~Shiomi}
\affiliation{College of Industrial Technology, Nihon University, 275-8576, Narashino, Chiba, Japan}
\author{M.~Subieta}
\affiliation{Instituto de Investigaciones F\'{i}sicas, Universidad Mayor de San Andres, 8635, La Paz, Bolivia}
\author{F.~Sugimoto}
\affiliation{Institute for Cosmic Ray Research, 277-8582, Kashiwa, Japan}
\author{N.~Tajima}
\affiliation{RIKEN, 351-0198, Wako, Saitama, Japan}
\author{W.~Takano}
\affiliation{Faculty of Engineering, Kanagawa University, 221-8686, Yokohoma, Kanagawa, Japan}
\author{Y.~Takeyama}
\affiliation{Faculty of Engineering, Yokohama National University, 240-8501, Yokohama, Kanagawa, Japan}
\author{M.~Takita}
\affiliation{Institute for Cosmic Ray Research, 277-8582, Kashiwa, Japan}
\author{N.~Tamaki}
\affiliation{Institute for Cosmic Ray Research, 277-8582, Kashiwa, Japan}
\author[0000-0001-9750-5440]{Y.~Tameda}
\affiliation{Faculty of Engineering, Osaka Electro-Communication University, 572-8530, Neyagawa, Osaka, Japan}
\author{K.~Tanaka}
\affiliation{Hiroshima City University, 731-394, Hiroshima, Hiroshima, Japan}
\author{R.~Ticona}
\affiliation{Instituto de Investigaciones F\'{i}sicas, Universidad Mayor de San Andres, 8635, La Paz, Bolivia}
\author[0000-0002-5581-0006]{H.~Tsuchiya}
\affiliation{Japan Atomic Energy Agency, 319-1195, Tokai-mura, Ibaraki, Japan}
\author{Y.~Tsunesada}
\affiliation{Graduate School of Science, Osaka Metropolitan University, 558-8585, Osaka, Osaka, Japan}
\affiliation{Nambu Yoichiro Institute for Theoretical and Experimental Physics, Osaka Metropolitan University, 558-8585, Osaka, Osaka, Japan}
\author{S.~Udo}
\affiliation{Faculty of Engineering, Kanagawa University, 221-8686, Yokohoma, Kanagawa, Japan}
\author{G.~Yamagishi}
\affiliation{Faculty of Engineering, Yokohama National University, 240-8501, Yokohama, Kanagawa, Japan}
\author{Y.~Yamanaka}
\affiliation{Institute for Cosmic Ray Research, 277-8582, Kashiwa, Japan}
\author{K.~Yamazaki}
\affiliation{College of Engineering, Chubu University, 487-8501, Kasugai, Aichi, Japan}
\author{Y.~Yokoe}
\affiliation{Institute for Cosmic Ray Research, 277-8582, Kashiwa, Japan}

\collaboration{99}{(The ALPACA Collaboration)}






\begin{abstract}
The Andes Large area PArticle detector for Cosmic ray and Astronomy (ALPACA) is a new air-shower array experiment under construction in the Bolivian Andes, and its prototype ALPAQUITA surface array has been operating since 2023 April.
In addition to the traditional $\ge$3-hit or $\ge$4-hit coincidences to trigger recording air-shower events, ALPAQUITA records the counting rates of the $\ge$1-hit and $\ge$2-hit events (Any1 and Any2, respectively).
We report a successful detection of a Forbush decrease occurred on 2024 May 10 caused by a passage of an interplanetary shock formed ahead of the Interplanetary Coronal Mass Ejection.
The amplitude detected in the Any1 rate is 4.26$\pm$0.33\% at the median primary rigidity of 76\,GV which is consistent with the observations with the worldwide neutron monitor and muon detector networks.
Under the assumption of a power-law rigidity spectrum, we renormalized the errors of the observed amplitude ($A_{obs}$) and fitted them as a function of the median primary rigidity ($R_{m}$) of each detector and observational method. 
The result $A_{obs} = (10.9\% \pm 0.9\%) \times (R_{m}/10\,GV)^{-0.55 \pm 0.07}$ exhibits a hard nature of this event.
Our non-detection in the Any2 rate decrease constrains the amplitude with a 2$\sigma$ upper limit to be 0.95\% at 960\,GV.
This marginally suggests an existence of a spectral softening between 100\,GV and 1000\,GV as also suggested by the Misato underground muon detector at 145\,GV.
Although a strong geomagnetic storm was observed during this period, we conclude it does not impact our results.
Our novel technique realizes a unique coverage to study the behavior of the Forbush decreases at the highest rigidity.
\end{abstract}

\section{Introduction}\label{sec:introduction}
Interplanetary Coronal Mass Ejections (ICMEs) associated with CMEs erupted from the Sun sometimes arrive at the earth. 
A magnetic flux rope connected to the Sun composes the ICME and it often forms a shock on its front. 
As the flux rope and the shock block the penetration of the galactic cosmic rays, a vicinity of low cosmic-ray density is formed and a passage of an ICME through the earth is observed as a sudden decrease of the cosmic-ray flux, which is known as ‘Forbush decrease (FD)’ (\cite{ref:forbush}).

FD is a unique phenomenon to study the propagation of cosmic rays through the turbulent magnetic field and shock. 
According to the review by \cite{ref:cane2000}, the amplitude of FDs follows the single power-law spectrum of the cosmic-ray rigidity. 
The variation of the spectrum index is reported from --0.4 to --1.2 and it is interesting to know the maximum rigidity where the FD amplitude may drop faster than the single power-law seen in the lower rigidity. 
The GRAPES-3 collaboration analyzed the rigidity dependence of the amplitude of 28 FDs observed with their muon telescope data  at various arrival directions together with the traditional measurements by the neutron monitors (NMs) to cover 10-100\,GV rigidity range (\cite{ref:grapes2013}).
They fitted the FD amplitude as a function of the median primary rigidity $R_{m}$ with a single power-law spectrum $R_{m}^{-\gamma}$ in two rigidity ranges. 
The NM data exhibit $\gamma$=0.65$\pm$0.05 in the rigidity range from 10.0 to 31.6\,GV while the muon data show a softening to $\gamma$=1.26$\pm$0.08 in R=64.4-92.0\,GV. 
On the other hand, \cite{ref:savic2019} reported observations of 4 FDs using the Belgrade CR station including an underground detector with 122\,GV median primary rigidity sensitivity. 
Combining with the NM data, their spectra are described with a single power ranging $\gamma$=0.58-0.86. 
While many FDs are observed beyond 100\,GV, its extension or softening/cutoff behavior at higher rigidity is still an open question.

FDs up to 30\,GV are accessed by NMs located at various altitudes and geomagnetic latitudes.
Higher rigidity is studied by muon detectors using their higher effective threshold energy to produce charged pions.
Directional selection or underground measurements of muons allow observations beyond 100\,GV median primary rigidity.
However, so far, technique to access FDs at multi 100\,GV or even TV is not discussed.
In this paper, a new technique to study FDs beyond 100\,GV using traditional air-shower array is proposed.
Though the air-shower observation requires simultaneous signals from at least 3 detectors to reconstruct the arrival direction and energy of the primary particle, the proposed method uses the counting rate of 2-hit coincidence to achieve lower threshold energy (rigidity).
The technique is applied in the new air-shower array ALPAQUITA located at 4,740\,m above sea level and achieved observations at sub-TV median primary rigidity.
The observation of the FD occurred on 2024 May 10 proves a potential of the new technique to study FDs between 100\,GV and 1000\,GV.

In this paper, we briefly summarize the extreme solar activity around 2024 May 10 in Sec.~\ref{sec:solar-activity}.
In Sec.~\ref{sec:alpaquita}, we describe the ALPAQUITA experiment and its counting mode operation.
Its corresponding rigidity is derived using a Monte Carlo simulation.
Section~\ref{sec:observations} reports the ALPAQUITA observation of the May 10 FD.
Section~\ref{sec:discussion} compares the ALPAQUITA results with the results obtained by the NMs and muon detectors, and Sec.\ref{sec:summary} summarizes the results. 

\section{Solar activity in 2024 May}\label{sec:solar-activity}
An extreme solar activity which induced low latitude auroras on the earth was observed in 2024 May.
A comprehensive review of this activity is summarized in \cite{ref:Hayakawa_2025}.
An FD was observed from May 10, 17\,UT by worldwide NM network.
Preceding to this FD, multiple X-ray class flares and associated CME emissions were observed, and the FD is thought to be a result of overlapping ICMEs.
At the beginning of the FD, the solar wind speed suddenly increased from 450\,km\,s$^{-1}$ to 700\,km\,s$^{-1}$ and the strength and direction of the interplanetary magnetic field became turbulent, suggesting an arrival of a shock front.

\begin{figure}[h]
\centering
\includegraphics[width=0.9\textwidth]{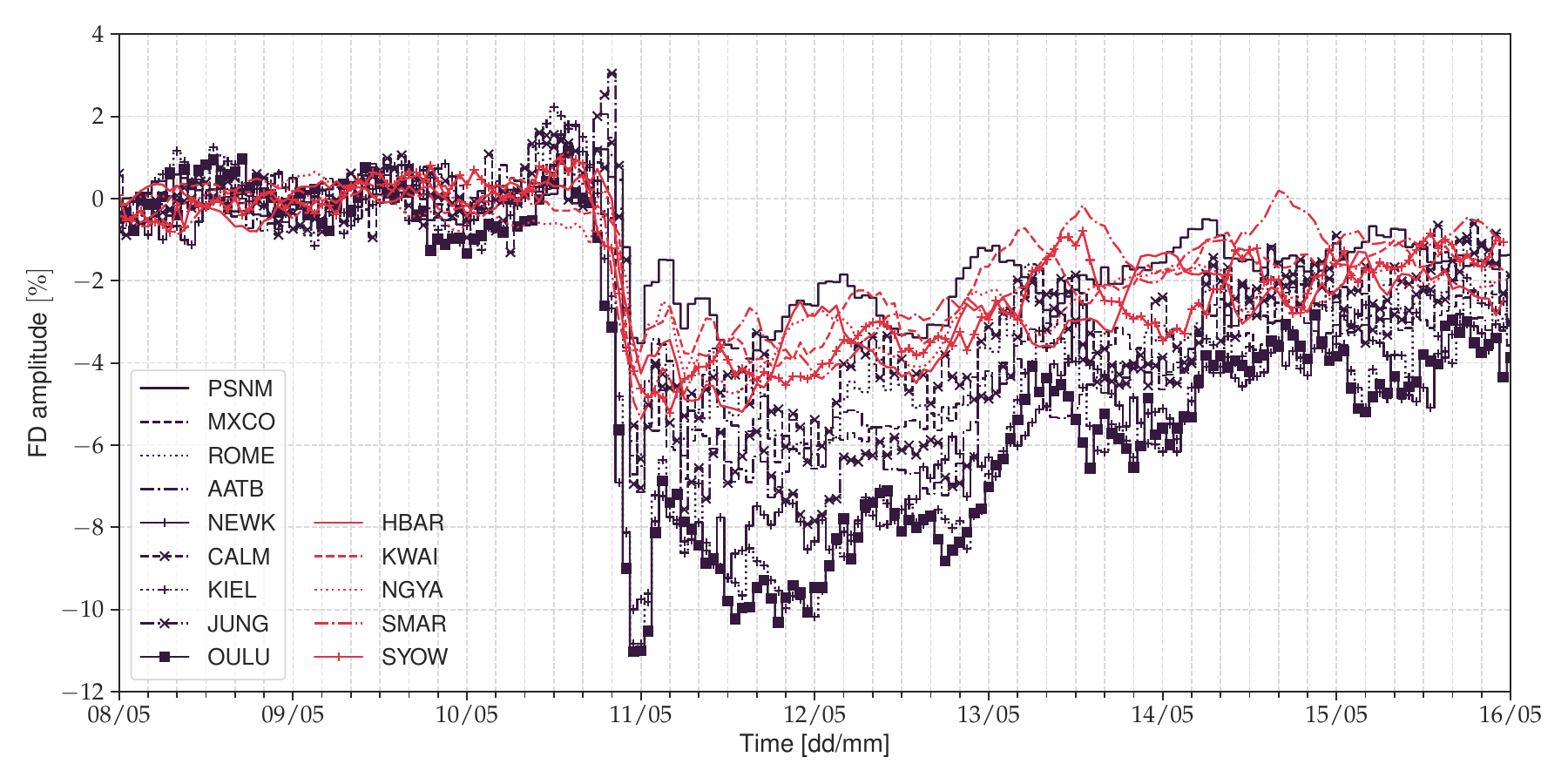}
\caption{Pressure corrected time profiles of the NM (black curves) and GMDN (red curves) relative counting rate during the FD started at around 17\,UT, May 10.}
\label{fig:NMs}
\end{figure}

Figure~\ref{fig:NMs} shows the relative 1-hour counting rates of NMs (Mexico city,  Princess Sirindhorn, Alma Ata, Newark, Castilla la Macha, Rome, Jungfraujoch, Kiel and Oulu stations) and the Global Muon Detector Network (GMDN) (five vertical channels of Hobart, Kuwait, Nagoya, S\~ao Martinho and Showa stations) data  (\cite{ref:GMDN}). The data from the NMs was obtained from the neutron monitor database and it is available online \footnote{\url{http://www.nmdb.eu/}}, while the archived GMDN data are available at the Shinshu University website \footnote{\url{http://hdl.handle.net/10091/0002001448}}.
The pressure effect is corrected and the base level is defined as an average counting rate during 48 hours in May 8 and 9.
Table~\ref{tab:stations} summarizes the instrument type and their name of the station or signal channel, detector area, altitude above sea level, median primary rigidity, average counting rate per hour and their observed amplitude as defined below.
The median primary rigidity is calculated following \cite{Munakata_2022} for a geomagnetically quiet period.
As seen in Fig.\ref{fig:NMs}, the intensity decreased to the minimum in the FD at around May 11, 0:00\,UT and rapidly increased by 3:00\,UT.
Then the counting rates exhibited another drop in May 12 and recovered gradually with the superposed diurnal variation.
The rapid increase at May 11, 3:00\,UT is recognized as a Ground Level Enhancement (GLE) \#74 in the GLE database \footnote{\url{https://gle.oulu.fi/}} (\cite{ref:gledatabase}) produced by the solar energetic particles associated with the X5.8 class solar flare occurred at May 11, 01:10\,UT.
To avoid the influence of this GLE, we define the amplitude of the FD at the first intensity minimum around May 11, 0:00\,UT.
Table~\ref{tab:stations} also includes the observed amplitude.
The error of the amplitude is calculated from the standard deviation of hourly counting rates used to calculate the base level average during 48 hours.
The calculated error is about 5-10\% of the observed amplitude.
Statistical uncertainty is negligibly small compared to this error.

\begin{table}[htp]
\caption{Detail of the observation sites and data used in this analysis. Columns (1) type of the instrument, (2) name of the station, (3) area of the detector, (4) altitude, (5) median primary rigidity, (6) average hourly count rate during 48 hours used to calculate the base level (see text), (7) observed amplitude of the FD. The area of ALPAQUITA is omitted because it requires a configuration of the air shower array seen in the main text.}
    \begin{center}
    \begin{tabular}{ccccccc}
Instrument & Station Name  & Detector area ($\mathrm{m}^{2}$) &  Altitude (m) & Median primary  & Counts per hour  & Observed\\
           & or Count Mode &                                  &               &  rigidity (GV)  &  $\times 10^5$     & Amplitude (\%)\\
\hline
NM & Jungfraujoch &  3.71 & 3475 & 13.5 & 11.7 &  8.04$\pm$0.47 \\
   & Oulu         & 11.14 &   15 & 14.9 &  3.6 & 10.85$\pm$0.56 \\
   & Kiel         & 22.27 &   54 & 15.2 &  5.6 & 10.77$\pm$0.59 \\
   & Newark       & 11.14 &   50 & 15.3 &  3.2 & 10.13$\pm$0.51 \\
   & AATB         & 22.27 & 3340 & 17.1 & 45.8 &  7.46$\pm$0.43 \\
   & Rome         & 24.75 &    0 & 19.5 &  4.4 &  6.35$\pm$0.49 \\
   & Mexico City  &  7.42 & 2274 & 20.4 &  7.9 &  7.19$\pm$0.31 \\
   & CALMA        & 18.56 &  708 & 20.4 &  2.5 &  6.39$\pm$0.49 \\
   & PSNM         & 22.27 & 2565 & 34.6 & 22.0 &  3.90$\pm$0.31 \\
\hline
GMDN & Hobart         & 16 &  65 &  53.1 & 14.6 & 4.89$\pm$0.30 \\
     & S\~ao Martinho & 28 & 488 &  54.3 & 23.6 & 5.15$\pm$0.38 \\
     & Showa          &  2 &  29 &  54.3 &  2.4 & 5.01$\pm$0.32 \\
     & Nagoya         & 36 &  77 &  58.4 & 29.1 & 4.69$\pm$0.40 \\
     & Kuwait         & 25 &  19 &  61.2 & 22.7 & 3.96$\pm$0.47 \\
\hline
Misato MD & Misato    & 16 & 735 & 145.0 &  2.7 & 1.94$\pm$0.22 \\
\hline
ALPAQUITA & Any1   & -- & 4740 &  76.0 & 2772 & 4.26$\pm$0.33        \\
          & Any2   & -- & 4740 & 960.0 &  169 & $<$0.95 (2$\sigma$UL)\\
\hline
    \end{tabular}
    \end{center}
  \label{tab:stations}
\end{table}%

\section{ALPAQUITA experiment}\label{sec:alpaquita}
The ALPAQUITA air-shower array has been operating since 2023 April in the plateau near the Mt. Chacaltaya in Bolivia (68$^{\circ}$08'\,W, 16$^{\circ}$23'\,S, 4,740\,m above sea level). ALPAQUITA is a prototype array of the Andes Large area PArticle detector for Cosmic ray physics and Astronomy (ALPACA) experiment. Combining the traditional surface detector array with the underground muon detectors established by the Tibet AS$\gamma$ experiment (\cite{ref:tibetMD}, \cite{ref:tibetCrab}), ALPACA and ALPAQUITA aim to explore the sub-PeV gamma-ray sky in the southern hemisphere for the first time (\cite{ref:alpaquita-kato}). The observations discussed in this paper are performed only using the ALPAQUITA surface detector array.

\begin{figure}[h]
\centering
\includegraphics[width=0.5\textwidth]{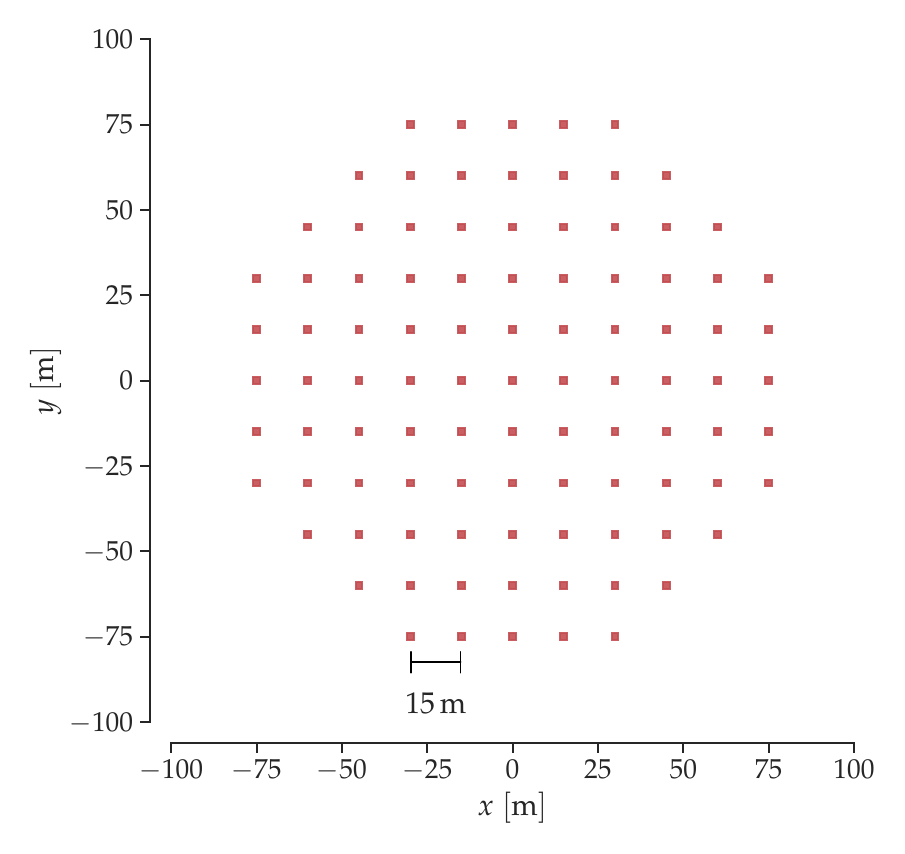}
\caption{Layout of the ALPAQUITA surface array. Each square represents a 1\,m$^{2}$ scintillating counter.}
\label{fig:alpaquita}
\end{figure}

The layout of the ALPAQUITA surface array is shown in Fig.\ref{fig:alpaquita}. Ninety-seven scintillating counters, of which 92 were in operation in 2024 May, are aligned on the grid of a 15\,m interval and cover a surface area of 18,450\,m$^{2}$. Each counter is composed of 1\,m$^{2}$ and 5\,cm thick plastic scintillator housed in a flat-top stainless box of 1\,mm thickness. A Lead plate of 5\,mm thickness is placed on the box to convert high-energy photons in the air-shower into electron-positron pairs. A photomultiplier tube of 2-inch diameter is placed at the bottom of the stainless box to measure scintillating light as electric signals. Air shower events are triggered when at least 4 (3 since 2024 June) counters exhibit signal above 0.4 minimum ionizing particle (4\,MeV equivalent deposit energy) within a 600\,ns time window. Accordingly, the charge and timing information of all hit counters are recorded. A report of the initial performance of the ALPAQUITA surface array is found in \cite{ref:ichepsako}.

In addition to the 3 (4) coincidence trigger scheme for the air-shower event recording, which we call Any3 (Any4) trigger, we implemented Any1 and Any2 triggers. The time window of the Any2 coincidence is the same as Any3 (4). The Any1 count rate is simply a sum of the single counting rates of each counter while the Any2 count rate is recorded to monitor the intensity variation of primary cosmic rays in an intermediate energy range between the energies responsible for the single particles and the air-shower events. Because we cannot reconstruct the arrival direction and energy of primary cosmic rays for the Any1 and Any2 events, only counting rates of them are recorded every 0.1 seconds. Typical counting rates of Any1 and Any2 are 77\,kHz and 4.7\,kHz, respectively, which are much higher than any detectors in Tab.\ref{tab:stations}.

\begin{figure}[h]
\centering
\includegraphics[width=0.6\textwidth]{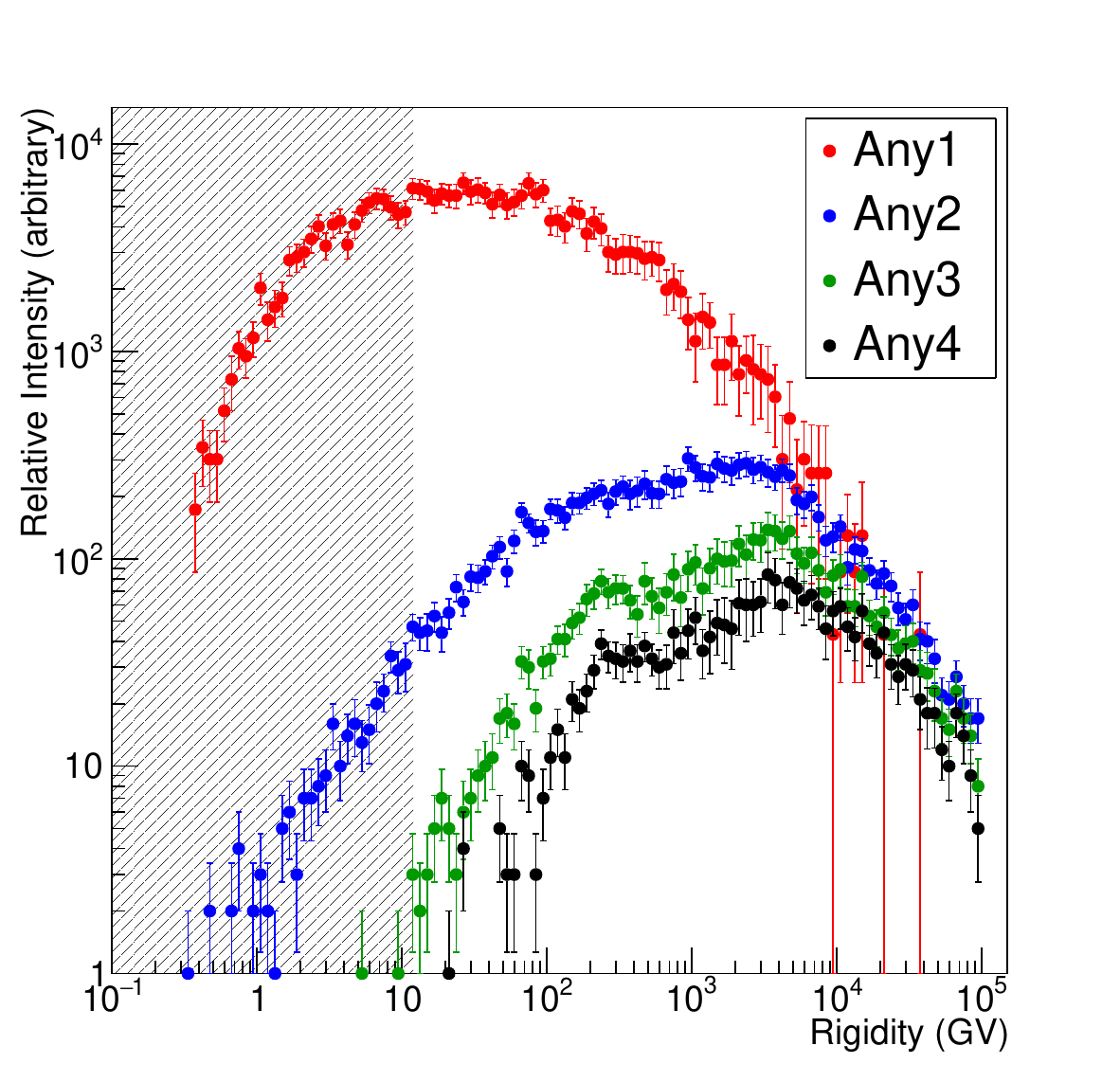}
\caption{Response functions of the ALPAQUITA Any1 (red), Any2 (blue), Any3 (green) and Any4 (black) counting modes. Rigidity range below 12\,GV, which is the vertical cutoff rigidity at the ALPAQUITA site, as indicated by shade is not included in the calculation of the median primary rigidities.}
\label{fig:response}
\end{figure}

We performed a Monte Carlo simulation study to determine the primary cosmic-ray rigidity responsible for each counting mode. Air shower events are generated using the CORSIKA~7.6400 package (\cite{ref:corsika}) where the FLUKA (\cite{BATTISTONI_FLUKA}) and QGSJET~II-04 (\cite{PRD_QGJII04}) hadronic interaction models are used below and above 80\,GeV laboratory frame energy, respectively. EGS4 (\cite{EGS4}) is used for the electromagnetic processes. A model of mass composition and mass-dependent energy spectra of the primary cosmic rays proposed by  \cite{ref:shibata-model} is used as an input at the top of atmosphere. To follow the single particle observations in Any1, the lowest input energy was set at 280\,MeV and the secondary particles are tracked down to the kinetic energy of 50\,MeV, 50\,MeV, 1\,MeV and 1\,MeV for hadrons (except $\pi^{0}$), muons, electrons and photons (and $\pi^{0}$), respectively. Particles arriving at the ALPAQUITA altitude are processed in the detector simulation implemented by the GEANT4~10.04.p02 simulation toolkit (\cite{ref:geant4}) and a custom-made simulation of the trigger electronics. Finally, the events passing the condition of Any1 to Any4 are counted. Fig.\ref{fig:response} shows the rigidity distribution of primary cosmic rays which passed the condition of each counting mode. The Any1 events have a peak rigidity at around a few 10\,GV. The Any2 events have a peak near 1000\,GV and it is close to the peaks of the air-shower triggers Any3 and Any4. 
It is also found that the response of each counting mode spreads over a wide rigidity range. 
The median primary rigidity of each counting mode is calculated from the response function. In this calculation the lowest rigidity was set at 12\,GV which is the vertical cutoff rigidity of the ALPAQUITA site. In Fig.\ref{fig:response} the rigidity range below 12\,GV is indicated as a hatched area. The resultant median primary rigidities are 76\,GV, 960\,GV, 2290\,GV and 3340\,GV for Any1, Any2, Any3 and Any4, respectively. The value of Any1 is similar to that of the muon detectors while the Any2 covers a unique rigidity not covered by the former studies of FDs.

\section{ALPAQUITA observation of 2024 May 10 Forbush Decrease}\label{sec:observations}
When the FD on May 10 was observed by the NMs and GMDN, the ALPAQUITA Any1 and Any2 counting rates also recorded decreases by 3.5\% and 5.2\%, respectively. 
It is surprising that the apparent amplitude of the Any2 decrease is larger than that of Any1 despite a higher median primary rigidity of Any2. 
We consider below the effect of the accidental coincidence where the uncorrelated single counts produce a fake count in Any2. 
Because the count rate of Any1 is $\sim$16 times larger than that in Any2, the accidental coincidence can make a significant effect on the Any2 count.

Let us define the counting rate of individual scintillation detector generated by the independent cosmic-ray incident as $r_{1}$. With $N$ detectors, the true Any1 counting rate is $R_{1,T}$=$r_{1} \times N$. Among these independent events, event pairs arriving in time within twice the pulse width of the logic signal, 2$\tau$, produce coincidence. This reduces the number of Any1 count while increases the fake Any2 count. The rate of this correction is $\binom{N}{2}\times 2\tau r_{1}^{2}$, where $\binom{N}{2}$ designates the number of 2 arbitrary choice from $N$. When the true Any2 rate by the high-rigidity cosmic-ray event is $R_{2,T}$, the observed Any1 and Any2 rates, $R_{1,O}$ and $R_{2,O}$, respectively, are described as,
\begin{equation}
R_{1,O} = R_{1,T} ~– \binom{N}{2}\times 2\tau r_{1}^{2} + R_{2,T}
\end{equation}
\begin{equation}
R_{2,O} = R_{2,T} + \binom{N}{2}\times 2\tau r_{1}^{2}. 
\end{equation}

Because the definition of Any-$n$ is the rate of $n$ or more hits,
$R_{2,T}$ is also counted in $R_{1,O}$.
The higher order effects from Any3 and Any4 are ignored in this correction.
Given $R_{1,O}$ and $R_{2,O}$ as the observed Any1 and Any2 rates, respectively, $N$ at the time of interest and $\tau$ from the electronics setup (effectively 550\,ns), we can obtain the $R_{1,T}$ and $R_{2,T}$. 
Though we need to solve a quadratic equation of $r_{1}$, considering a relation $R_{1,T} \sim R_{1,O}$, the solution is uniquely determined. 
We found that the fraction of $R_{2,T}$ to $R_{2,O}$ is about 20\% while the difference between $R_{1,T}$ and $R_{1,O}$ is less than 1\%.

\begin{figure}[h]
\centering
\includegraphics[width=0.8\textwidth]{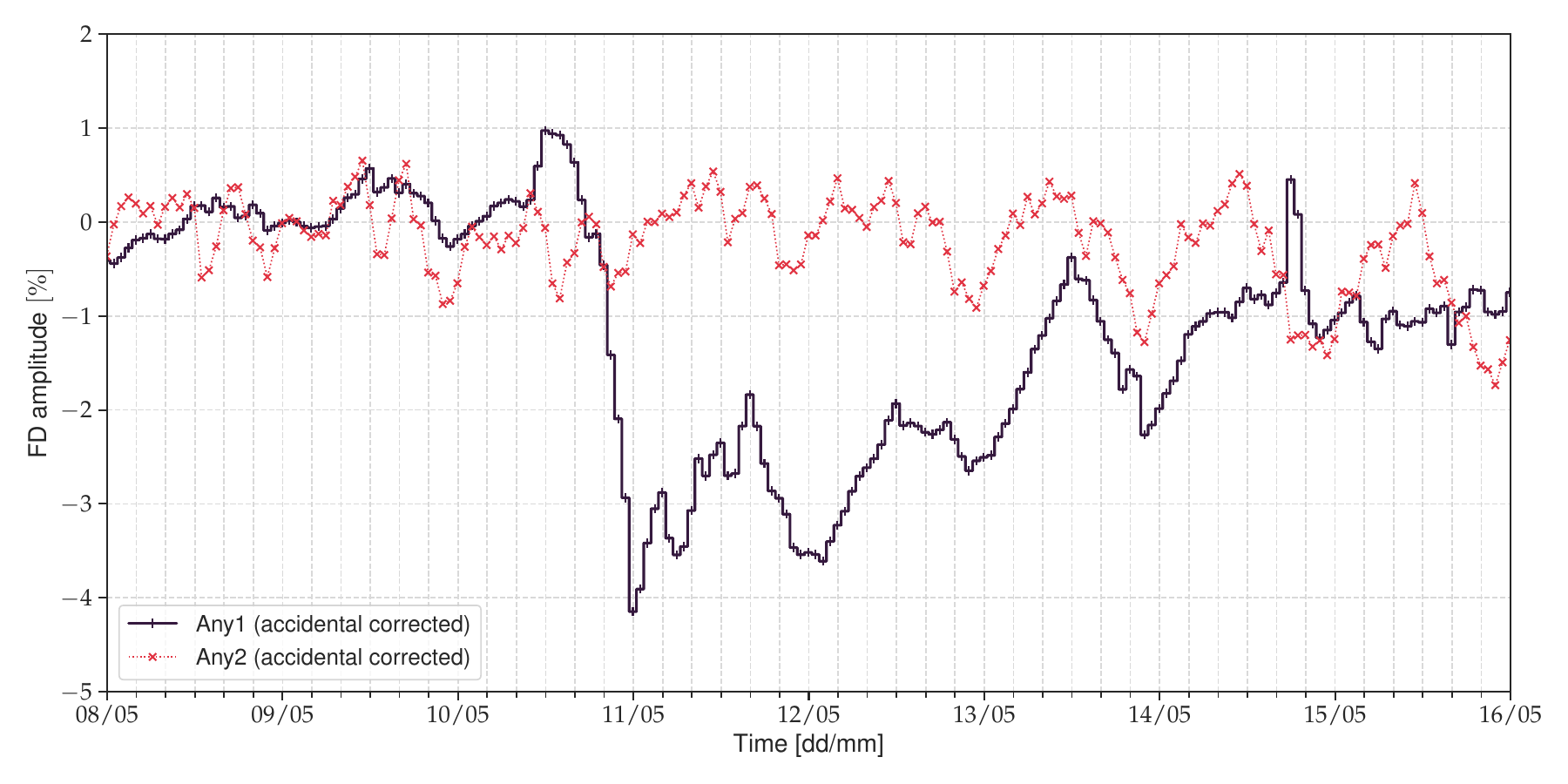}
\caption{Time profiles of the ALPAQUITA Any1 (black) and Any2 (red crosses) relative counting rates after the accidental coincidence and pressure corrections.}
\label{fig:any1any2}
\end{figure}

Figure~\ref{fig:any1any2} shows the corrected counting rates of Any1 and Any2 ($R_{1,T}$ and $R_{2,T}$) around the FD event after applying the corrections for the atmospheric pressure effect (barometric coefficients are -0.47\%/hPa and -0.78\%/hPa for Any1 and Any2, respectively). Though the decrease in Any1 remains, the Any2 counting rate no more indicates significant decrease.
The amplitude and its error of the Any1 decrease, 4.26$\pm$0.33\%, are estimated in the same manner discussed in Sec.\ref{sec:solar-activity} and summarized in Tab.\ref{tab:stations}.
We defined the 2$\sigma$ upper limit of the Any2 amplitude to be 0.95\%, as a sum of the observed decrease (0.23\%) at the moment of the FD peak and twice the standard deviation of the Any2 counting rate in May 8 and 9 (0.36\%).
It is also summarized in Tab.\ref{tab:stations}.

\section{Discussions}\label{sec:discussion}
\begin{figure}[h]
\centering
\includegraphics[width=0.8\textwidth]{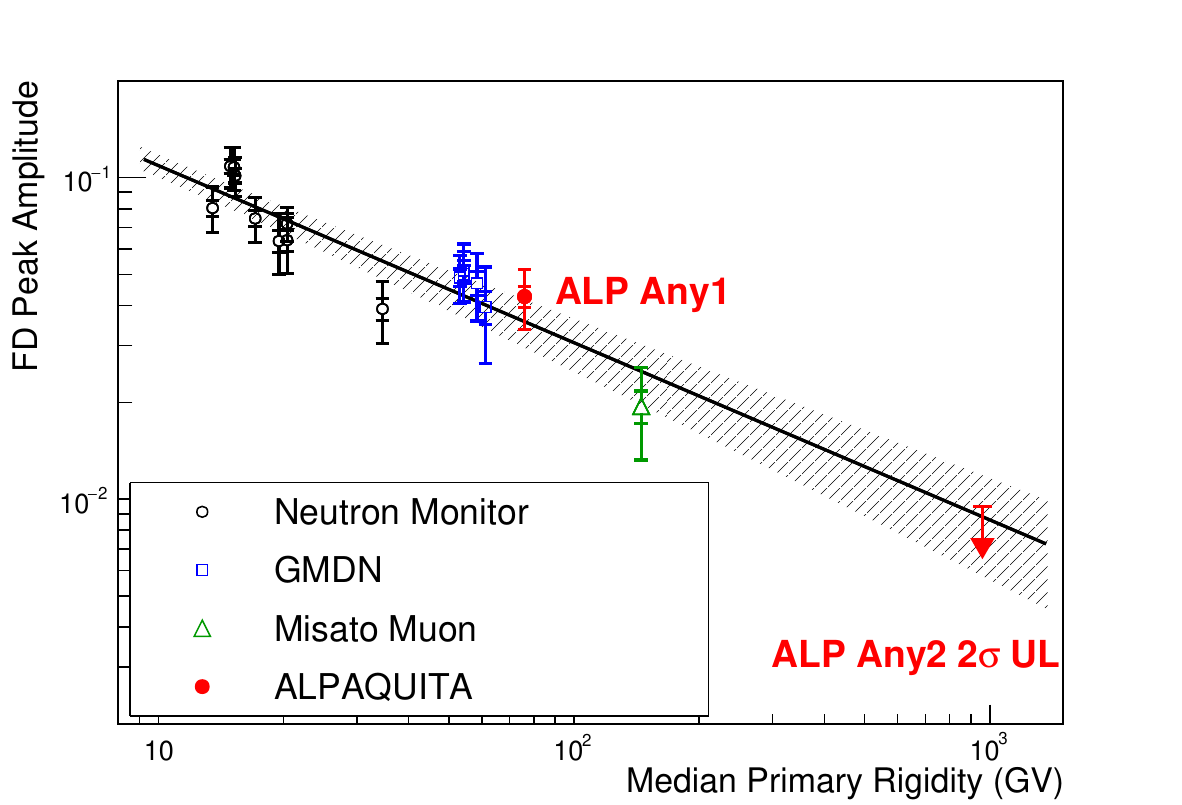}
\caption{Median primary rigidity and amplitude of the FD event on May 10, 2024. Black open circles, blue open squares, green open triangle and red filled circle show the data of NMs, GMDN, Misato muon detector and ALPAQUITA Any1, respectively. The red arrow shows the upper limit determined by the ALPAQUITA Any2. Narrower error bars in each plot are derived from the standard deviation of each station while the wider errors are defined to obtain $\chi^2$/dof=1 in the power law fitting shown with a black line and hatched area indicating 1$\sigma$ error band.}
\label{fig:spectrum}
\end{figure}

Figure~\ref{fig:spectrum} summarizes a relation between the median primary rigidity ($R_{m}$) and the observed FD amplitude of the various sites and observational methods.
Two error ranges are assigned to each data point, where the narrower error is the one defined in Sec.\ref{sec:solar-activity} and summarized in Tab.\ref{tab:stations} while the wider error is defined below. 
A power law fitting without including the ALPAQUITA Any2 upper limit point and using the narrow errors gives a best fit function of observed amplitude ($A_{obs}$) is $A_{obs} = (10.9\% \pm 0.3\%) \times (R_{m}/10\,GV)^{-0.55\pm0.03}$ with $\chi^2$=108.1 (14 degrees of freedom).
Apparently, the large $\chi^{2}$ does not support this simple fitting.
As seen in Fig.~\ref{fig:spectrum}, the amplitude of many data points largely deviate from the neighbor points with respect to the narrower error sizes even if they have similar median primary rigidities.
This is possible when the FD has a significant diurnal variation superposed due to the anisotropy and the FD amplitude is determined not only by the median primary rigidity but also by the detector's viewing longitude.
Looking at Fig.\ref{fig:NMs} and Fig.\ref{fig:any1any2}, a clear diurnal modulation of the amplitude is found after May 12.
Slight differences between the maximum phases of diurnal variation observed by different stations support an existence of the influence of strong anisotropy.
Though a sophisticated technique to resolve this diurnal anisotropy is developed by \cite{Munakata_2022}, application of such correction is not in the scope of this paper.
Instead of correcting the anisotropy effect, we simply rescaled the errors by multiplying a factor 2.78 ($\sqrt{108.1/14}$) to make $\chi^2$/dof equal to unity.
The rescaled errors are shown as the wider error ranges in Fig.\ref{fig:spectrum}.
This gives a conservative error estimate of the fitting parameters under an assumption of power law nature of the amplitude.
The new fitting shown in Fig.~\ref{fig:spectrum} with a 1$\sigma$ error band gives $A_{obs} = (10.9\% \pm 0.9\%) \times (R_{m}/10\,GV)^{-0.55\pm0.07}$.

We discuss the validity of this artificial scaling factor 2.78 in terms of anisotropy.
\cite{Munakata_2024} reported the amplitudes of diurnal anisotropy during two FD events observed in 2012 to be 1.5\% and 2\% at 15\,GV.
From Tab.\ref{tab:stations}, the error of the amplitude at 15\,GV is about 0.5\%.
Scaling this by 2.78 results in an error of 1.4\%, which is consistent with the diurnal anisotropy amplitude reported in the previous study. 
Recently, \cite{ABUNINA20257578} reported the anisotropy of the 2024 May 10 FD event and concluded the north-south and equatorial anisotropies were 3.1\% and 1.9\%, respectively.
Because these values are the moduli of the maximum-to-minimum variations,  
a quadratic sum of half of them, 1.8\%, can be compared to 1.4\%, and they are in a reasonable agreement.
We note that anisotropies during large FDs are frequently reported (eg. \cite{HOFER_FLUCKIGER_2000} \cite{MISHEV20244160}), and \cite{ABUNINA20257578} point out that the anisotropy of the 2024 May 10 event is the smallest compared with the other large ($>$10\% amplitude) FDs.  

The power index of 0.55 is situated in the hard side of the range reviewed in \cite{ref:cane2000}, which allows us to discuss the high-rigidity behavior of this event.
The measurement by the ALPAQUITA Any1 at 76\,GV (red filled circle) well aligns on the power-law.
Comparing with the GMDN data at slightly lower rigidity (blue open squares), there is no indication of softening.
On the other hand, the observation with the Misato underground muon detector at 145\,GV (green open triangle), indicates a spectral softening.
Beyond a wide gap from the other points, the 2$\sigma$ upper limit amplitude of the ALPAQUITA Any2 at 960\,GV is located on the extrapolation of the low rigidity measurements.
Though it is marginal in this event, the upper limit of the ALPAQUITA Any2 constrains a simple extrapolation from the low energy behavior.
This result demonstrates the ability of new technique to study the high-rigidity behavior of FD events.

We note that the observed FD amplitude tends to become larger than that in space with wider spread of the response function.
In general, the response function of a muon detector spreads wider than that of a NM.
Also, as we see in Fig.\ref{fig:response}, the response function of ALPAQUITA Any1 spreads in a wide rigidity range.
These facts possibly make the amplitude at higher rigidity in this study systematically larger and consequently a harder rigidity dependence.
When we discuss the rigidity dependence of FD amplitude in space instead of the dependence of the observed amplitude, we must take this point into account.

Possible impacts of the geomagnetic storm observed during this period are discussed.
\cite{ref:Hayakawa_2025} reported that the Dst index observed at 2\,UT on May 11 ranks this event as the sixth-largest geomagnetic storm since 1957.
In general, a geomagnetic storm reduces the geomagnetic cutoff rigidity and consequently increases the observed FD amplitude.
Because the median primary rigidities summarized in Tab.\ref{tab:stations} are given for the magnetically quiet period, a possible reduction of the median primary rigidity moves the point in Fig.~\ref{fig:spectrum} leftward or a possible correction of this effect moves the point downward.
\cite{ABUNINA20257578} estimated this so-called magnetospheric effect on the NM observation on May 10 at 10\,GV.
They concluded that within the observed 15.7\% amplitude, a 4\% amplitude is a contribution from the magnetospheric effect.
\cite{Munakata_2018} discussed that the reduction of the cutoff rigidity is small enough for the observation sites with the quiet-time cutoff rigidity above 10\,GV.
In addition, the detectors counting high-energy particles by requiring multi-channel coincidence, such as GMDN and ALPAQUITA Any2 or by operating at underground such as the Misato muon detector have response to primary particles only above 10\,GV.
In such detectors, even if their site has a low cutoff rigidity, they are not sensitive to geomagnetic storms.
As discussed in Sec.\ref{fig:alpaquita}, the quiet-time cutoff rigidity of the ALPACA site is 12\,GV.
We calculated the cutoff rigidity during the May 11 geomagnetic storm using the OTSO tool (\cite{JGR_OTSO}) using the parameters given in \cite{papaioannou2025highenergyprotonsgroundlevel} at the time of GLE\,74 and concluded that the reduction is 0.7\,GV.
It reduces the median primary rigidity by 4\,GV with respect to the quiet-time value of 76\,GV and thus has a negligible impact on our discussions.
In conclusion, the magnetospheric effect can move only some NM data points in Fig.~\ref{fig:spectrum} downward but does not impact the other points.
It does not change or even strengthens our conclusions regarding the hard nature of this event and the importance of the ALPAQUITA Any2 observation.

\section{Summary}\label{sec:summary}
We implemented the counting modes of Any-$n$ coincidence in the newly constructed ALPAQUITA air-shower array.
Though the Any1 and Any2 events are not useful for the air-shower reconstruction, their counting rates, especially Any2, have a unique rigidity coverage to monitor the variability of the cosmic-ray intensity.
We successfully detected the FD occurred from 17:00~UT on 2024 May 10 together with the NM and GMDN networks.
Our Any1 measurement with the 76\,GV median primary rigidity aligns on the power-law rigidity dependence consistently with the other measurements.
Though the Any2 count does not detect any significant decrease at 960\,GV, its amplitude upper limit marginally suggests an existence of a spectral softening between 100\,GV and 1000\,GV, consistent with another indication by the Misato underground muon detector at 145\,GV.

It is clear that new measurements sensitive at a few 100\,GV are necessary to clearly identify the rigidity limit of FDs.
Using the air-shower array technique implemented in this study, smaller detector interval is a promising direction.
ALPAQUITA is now preparing a small but denser area with a 7.5\,m detector interval.
In addition, to avoid a significant contamination from the accidental coincidence of independent single particles, An Any2 trigger with segmented areas are also being investigated.
These upgrades will enable the identification of the highest-rigidity behavior of FDs, hence the research of cosmic-ray penetration into the turbulent magnetic field and shocks.  

\vskip 5mm
{\bf Acknowlegdements}
We acknowledge the NMDB database \url{www.nmdb.eu}, founded under the European Union's FP7 programme (contract no. 213007) for providing data and to the PIs of each individual neutron monitor. 
We also acknowledge the GMDN collaboration and the Shinshu University for providing the data observed with the GMDN and the Misato underground muon detector (\url{http://hdl.handle.net/10091/0002001448}). 
The ALPACA project is supported by the Japan Society for the Promotion
of Science (JSPS) through Grants-in-Aid for Scientific Research (A)
24H00220, Scientific Research (B) 19H01922, Scientific Research (B)
20H01920, Scientific Research (S) 20H05640, Scientific Research (B)
20H01234, Scientific Research (B) 22H01234, Scientific Research (C)
22K03660, Grant-in-Aid for Challenging Research (Pioneering) 24K21200 and Specially Promoted Research 22H04912, the LeoAtrox
supercomputer located at the facilities of the Centro de An\'{a}lisis
de Datos (CADS), CGSAIT, Universidad de Guadalajara, M\'{e}xico, and
by the joint research program of the Institute for Cosmic Ray Research
(ICRR), The University of Tokyo. 
Y.~Katayose is also supported by JSPS Open Partnership joint Research projects F2018, F2019. 
K.~Kawata is supported by the Toray Science Foundation.  
E.~de~la~Fuente thanks financial support from Inter-University Research Program of the Institute for Cosmic Ray Research, The University of Tokyo, grant 2023i-F-005. 
I.~Toledano-Juarez acknowledges support from CONACyT, M\'{e}xico; grant 754851.
S.~Kato acknowledges support from the Agence National de la Recherche (ANR), project ANR-23-CPJ1-0103-01.
Y.~Hayashi is supported by JST SPRING, Japan, Grant Number JPMJSP2144 (Shinshu University).

\bibliography{sample631}{}
\bibliographystyle{aasjournal}



\end{document}